# Photonic crystals for enhanced light extraction from 2D materials

**Yasir J. Noori, Yameng Cao, Jonathan Roberts, Christopher Woodhead, Ramon Bernardo-Gavito, Peter Tovee and Robert J. Young**

Department of Physics, Lancaster University, Lancaster, LA1 4YB, UK

E-mail: r.j.young@lancaster.ac.uk

**Abstract**

We propose a scheme for coupling 2D materials to an engineered cavity based on a defective rod type photonic crystal lattice. We show results from numerical modelling of the suggested cavity design, and propose using the height profile of a 2D material transferred on top of the cavity to maximise coupling between exciton recombination and the cavity mode. The photonic structure plays a key role in enhancing the launch efficiency, by improving the directionality of the emitted light to better couple it into an external optical system. When using the photonic structure, we measured an increase in the extraction ratio by a factor of 3.4. We investigated the variations in the flux spectrum when the radius of the rods is modified, and when the 2D material droops to a range of different heights within the cavity. We found an optimum enhancement when the rods have a radius equal to 0.165 times the lattice constant, this enhancement reduces when the radius is reduced or increased. Finally, we discuss the possible use of solid immersion lenses, in combination with our photonic structure, to further enhance the launch efficiency and to improve vertical confinement of the cavity mode.

**Introduction**

Solid-state lighting has enabled a vast range of applications, making it one of the most important technologies of the 21$^{st}$ century, recently recognised by the Nobel Prize in Physics in 2014 [1]. Careful engineering of solid-state devices has enabled the creation of quantum light sources, capable of producing single and entangled photons [2, 3]. Quantum light sources provide a variety of unique applications such as securely share information between two parties using quantum key distribution (QKD) protocols [4]. QKD is a mature technology with a number of commercial systems available on the market today. Unfortunately these are bulky, expensive and generate their photons statistically by attenuating laser pulses, leaving them open to attacks [5]. Currently, there is a large drive to develop full QKD systems at the microscopic scale [6], incorporating the aforementioned quantum light sources as their qubit generators. Other nascent applications of quantum light include beating the classical diffraction limit in quantum imaging [7, 8], reducing cell damage in the microscopy of biological systems [9], random number generators [10], high resolution metrology [11] and linear all-optical quantum computing [12, 13].

Existing implementations of single photon sources, using atoms [14, 15], organic molecules [16], nitrogen vacancies in diamond [17] or semiconductor quantum dots [18, 19] often require precise control of laboratory conditions to operate, leading to unsolved challenges prohibiting wide spread adoption of the technology. Recently, considerable interest has been given to single photon sources based on two-dimensional (2D) direct-gap quantum emitters, such as monolayer transition metal dichalcogenides (TMDC). For example, sources based on molybdenum disulphide ($MoS_2$) and tungsten diselenide ($WSe_2$) have been demonstrated [20, 21], and they are significant as they are compatible with conventional silicon photodetectors. Furthermore, due to the large spin-orbit coupling [22], it has been shown that pure spin states can be electrically controlled in these monolayers, with them being protected against decoherence by the spin-split valence band. Another desirable attribute of 2D quantum emitters, and possibly the most interesting, is their ability to be reliably transferred onto different substrates [23], enabling integration into on-chip quantum photonic circuits. The current bottleneck for the application of this technology is the low optical absorption and emission of freestanding monolayers of a few percent, which severely limits the efficiency of light collection and extraction from potential optoelectronic implementations [24]. Therefore, a suspended monolayer emitter coupled to a photonic crystal resonant cavity [25], which can enhance quantum light collection and suppresses absorption by the substrate, constitutes a promising system to realise a fully integrated single photon source.

Monolayer-cavity coupling has been demonstrated recently using an air-bridge photonic crystal. This method in principle offers advantages including contact fabrication and minimised vertical mode loss, via the refractive index differential [26]. However, there is a reduction in the light-matter coupling strength since the geometry prevents spatial alignment between the emitter above the bridge and the cavity mode within the bridge slab. Moreover, curvature in the monolayer sheets, may further suppress the coupling between the photonic mode and the emitter. In the present work, we consider a monolayer embedded in a photonic crystal that consists of silicon rods arranged in a triangular lattice with a missing rod off the lattice forming an H1 resonant cavity, as shown in figure 1a. The important figure of merit here is the light extraction ratio $\eta_e$, defined as:

$$\boldsymbol{\eta_e = P_{cav}/P_0} \tag{1}$$

where $P_{cav}$ corresponds to the power collected from a monolayer coupled to the photonic cavity, and $P_0$ corresponds to the reference power measured for a monolayer exfoliated on top of a silicon substrate. It has been also been shown that monolayers suspended on hollowed structures can promote the creation of localised excitons due to lateral stress [27]. Interestingly, the spatially distorted region of the monolayer between rods leads to improved spatial coupling between localised excitons and the cavity-mode's antinode. This makes the structure a potential candidate for engineered and coupled defects in 2D materials as a source for quantum light applications.

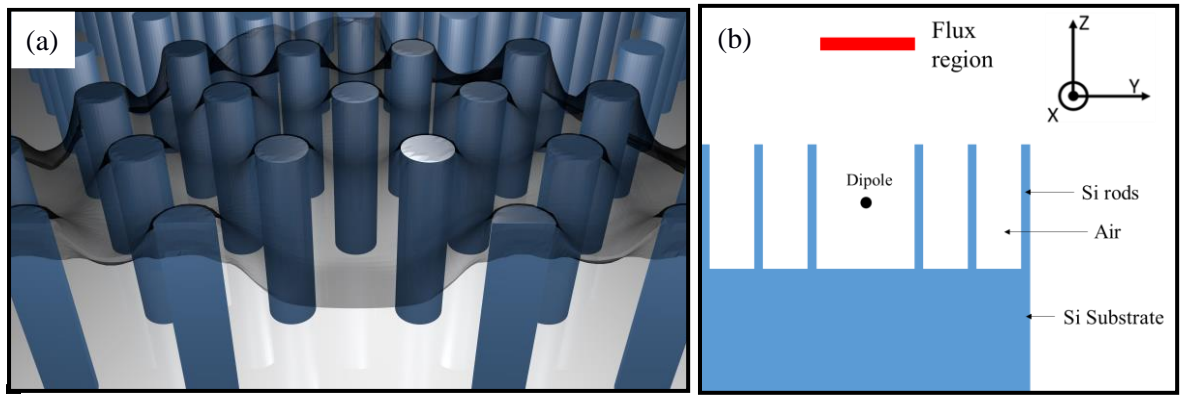

Figure 1: (a) Cross sectional illustration of a silicon rod photonic crystal cavity with a monolayer transferred on top of it (b) Cross sectional schematic diagram of the 3D simulated cavity.

**Methods**

The cavity design we propose consists of a triangular array of rods with a refractive index of 3.9, corresponding to the refractive index of silicon at the MoS$_2$ monolayer emission wavelength of approximately 660 nm. While this structure can be scaled and adjusted to match any 2D material emitter wavelength, we consider the MoS$_2$ emission wavelength since it is a very promising quantum light source in optoelectronic applications. The rods were chosen to be surrounded by air for optimum index contrast. By missing a rod from the centre of the lattice, an H1-defect cavity is formed and is used for this study. This arrangement is illustrated in figure 2a. Three-dimensional finite difference time domain (FDTD) simulations of the dipole cavity modes of this H1 rod-type photonic crystal cavity were performed using open-source software written by Oskooi et al. [28]. In all the simulations performed, perfectly matched layers were incorporated at the boundaries of the simulation domain to avoid unnecessary reflection of light. The radii of the lattice dielectric rods was varied between $0.155\alpha$ and $0.180\alpha$ (where $\alpha$ is the lattice constant) to find the optimum point in which the Q-factor and the enhancement in light extraction efficiency is calculated. Choosing the rod heights is crucial in this situation to maximise the cavity's Q-factor and the lattice band gap size [29]. The rod height should not be too small as this can cause the mode to form in air (above the structure), where the reduction in the interface between the spatial mode and the rods diminishes the Q-factor. On the other hand, they should not be too large as this can allow higher order modes and propagating modes to occupy the structure, and also requires high resolution lithography. Previous work has shown that the maximum gap size for a square rod-type photonic crystals is calculated for rod heights of approximately $2.3\alpha$ [29]. In our work, we found that this also applies to triangular lattices. Therefore the rods' height was fixed to $2.3\alpha$ in all simulation runs.

**Results and Discussion**

Using 2D plane wave expansion (PWE) simulation methods with the triangular lattice photonic crystal shown in Figure 2a, we obtained a photonic band gap for transverse electric (TE)-like modes at normalised wavelengths between $\lambda=1.07\text{-}1.72\alpha$. Subsequently, a defect was created by missing a single rod from the lattice in a location surrounded by 6 or more lattice points. For rods with radii of $0.165\alpha$, this creates a localised state at a normalised wavelength, $1.13\alpha$, resulting in a light trapping cavity.

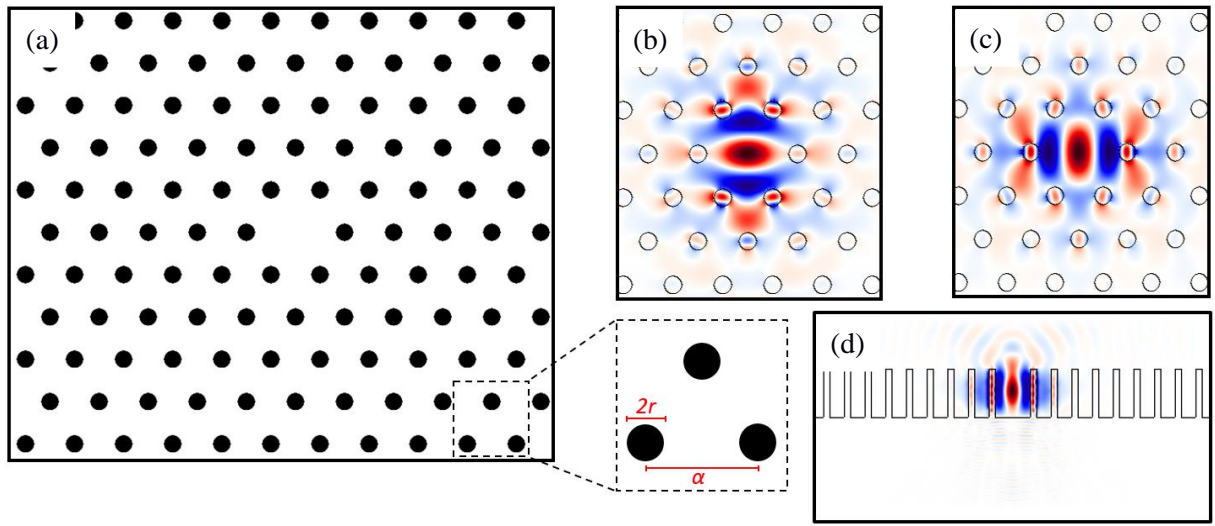

Figure 2: (a) An x-y slice of the simulated photonic crystal cavity structure through the rod structure. 3D FDTD simulation of the photonic cavity mode showing time slice images of the confined (b) $E_x$ (c) $E_y$ and (d) cross-section of the $E_y$ field components within the micro-cavity.

The simulation was run initially with an $E_x$ polarised source placed exactly in the centre of the cavity. A broadband source was setup for initial identification of the confined mode wavelength. An FDTD time slice of the spatial distribution of the cavity mode in the vicinity of the defect for the $E_x$ mode is shown in figure 2b. The second component of the TE mode in the plane of the photonic crystal lattice, the $E_y$ mode, was simulated and the designed cavity was shown to have a confined mode as shown in figure 2c. It is clear from the figure that the shape of the defect localised mode for the H1 cavity exhibits the character of the underlying hexagonal lattice. Results for the resonance wavelength of the cavity and its Q-factor were obtained using the harmonic inversion solving technique [30]. The Q-factor was measured to be 110 when the full photonic crystal structure was simulated using 3D FDTD. This relatively low Q-factor is limited by the lack of an air-semiconductor boundary in the vertical direction.

Modelling photonic crystal cavities using periodic dielectric rods has been reported previously [31]. A square lattice structure of dielectric rods with radii $0.2\alpha$ was used to realise a transverse magnetic mode cavity, producing a cavity mode wavelength at $\lambda = 2.6\alpha$. The main drawback that this design suffers from is the small radius requirement of the dielectric rods for a predefined cavity mode wavelength. For example, enhancing light from $MoS_2$ TMDC monolayers, at a wavelength of 660 nm, would require a lattice constant to be as small as 250 nm and the rods radius to be 50 nm. This renders the fabrication of such cavities a significant challenge for mode wavelengths in the visible regime, even with state of the art e-beam lithography techniques and highly tuned anisotropic etching recipes. Using the pre-mentioned hexagonal array design of dielectric rods from this work, the lattice constant and rods' radii are approximately 580 nm and 85 nm respectively, alleviating some of the fabrication challenges. The practical properties of this design make it more attractive to study in the application of monolayers for optoelectronic devices. In addition, the large diameter size of the photonic crystal rods, which distinguishes our design, allows easier transfer of 2D materials on top of the rod structures. This is due to the improved surface contact between the top surface of the rods and the 2D flakes, leading to higher adhesion.

To measure the improved extraction efficiency of light we mimic an objective lens with a numerical aperture of 0.65 with a flux region of area $23\alpha^2$ setup above the cavity, collecting vertically radiated light at a height of $3.5\alpha$. Using this setup we studied the collection efficiency enhancement due to a rod-type photonic crystal cavity. The comparison was made between a source within the 3D modelled cavity and a source above a silicon slab, where the dipole represents a 2D flake suspended on the structure as shown in figure 1b, and a flake exfoliated on a silicon substrate respectively.

Modelling the collected flux as a function of different photonic crystal rod radii, r, was carried out in order to investigate how non-uniformities in the structure fabrication could affect the enhancement in the collected light. Figure 3a illustrates the collected flux spectrum as the rods' radius is changed from $0.155\alpha$ to $0.180\alpha$. The results show that there was no enhancement in the collected light when r= $0.160\alpha$ or below, this is due to the lack of a lateral confinement as the radius of the rods is reduced. Enhancement in the collected flux peaks abruptly at $0.165\alpha$ and reduces to different levels as the radius is increased to $0.180\alpha$. We noticed that reducing the radius red-shifts the cavity's resonance wavelength. This is anticipated as the relationships between cell diameters and the resonance wavelength was modelled in many photonic crystal cavity studies [32].

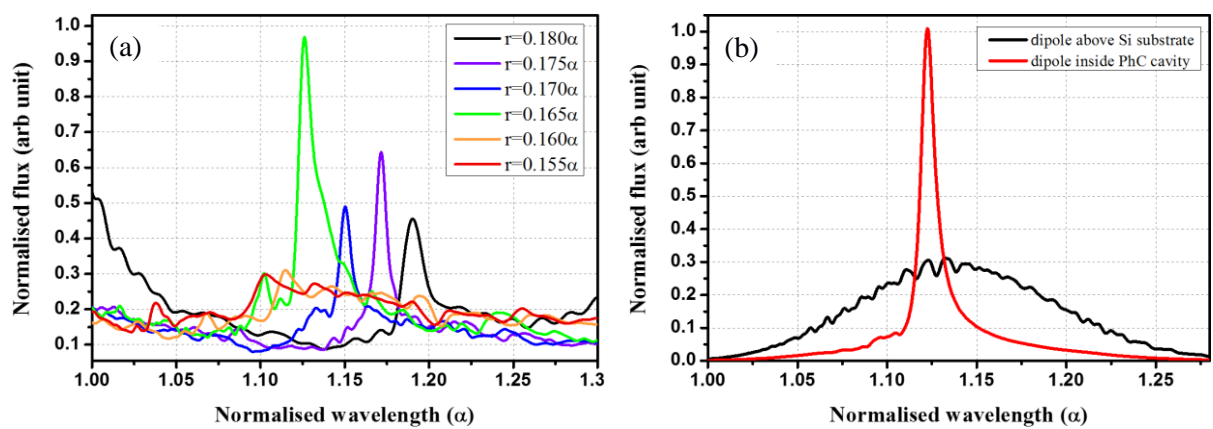

Figure 3: (a) The enhancement spectra for rods with a radius ranging from $0.155\alpha$ to $0.180\alpha$ (b) the normalised flux spectrum for a crystal with r=$0.165\alpha$ compared with the emission from a source on a bulk substrate.

Figure 3b plots a comparison between the normalised flux for an emitter inside a rod-type cavity of r=$0.165\alpha$ and for an emitter placed directly above a flat substrate. When the source is placed directly on top of the substrate a Gaussian curve is obtained which corresponds to the input dipole source emission wavelength and bandwidth as shown in the black curve in figure 3b. On the other hand, placing the dipole within the vicinity of the cavity results in an enhancement in the recorded flux, as shown by the red curve. This is attributed to radiation modes from the cavity leaking vertically toward the flux region. The light extraction ratio, $\eta_e$, from equation 1, was found to be approximately 3.4 for the case when the rods radius is $0.165\alpha$.

While stress due to the suspension of 2D monolayer flakes can promote the creation of light emitting defects [27], the exact position of the defect in the vertical direction inside the cavity cannot be easily anticipated without setting up simulations using finite element methods. Figure 4 shows simulation results of different flux spectra collected as the position of the dipole emitter is changed along the z-axis from the bottom of the cavity to the top. Here the rod height is chosen to be twice the cavity mode wavelength. As expected from figure 4, the maximum enhancement is achieved when the source is placed at a vertical position of half the height of the rods. In other words, at the centre point of the cavity, which corresponds to one wavelength above the substrate. Enhancement reduces dramatically when the source is near the top of the rods. This is because light escapes the cavity easily at this position. Whilst when the source is very close to the substrate as shown, the majority of the light emitted by the dipole is reflected at the air substrate interface. The reflected light from the bottom of the cavity gets enough time to interfere with the cavity and radiate away as a cavity mode.

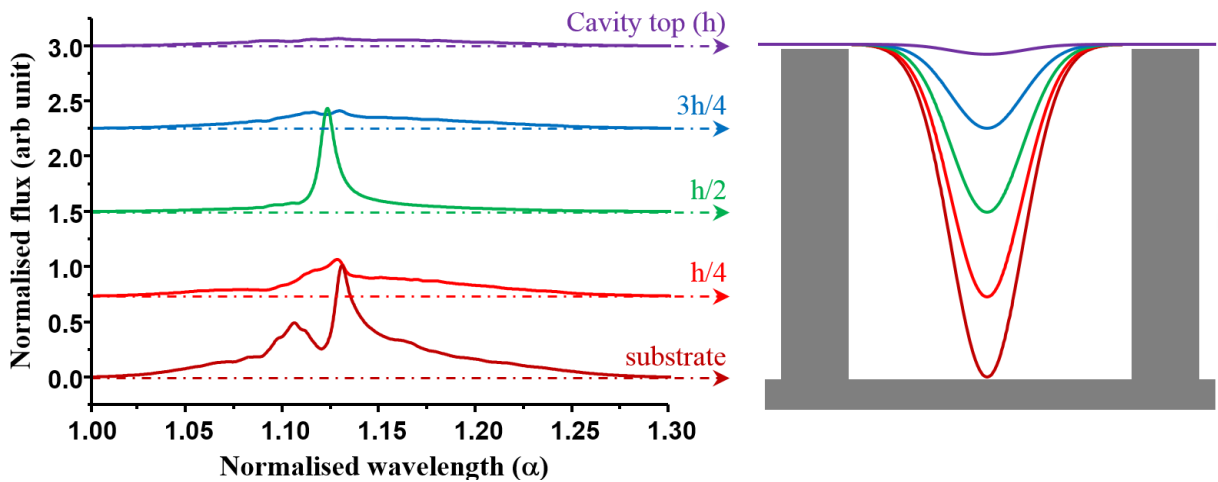

Figure 4: A comparison of the flux spectra for emission from the cavity as the dipole position inside the cavity is varied along the z axis. The drawing on the right illustrates the degrees of bowing of the 2D material from the surface of the structure into the cavity for each of the cases shown on the left.

The collection efficiency of emission from the cavity could be further improved by adding a solid immersion lens (SIL) on top of the photonic crystal structure. Glass SILs can be reliably positioned with micron-scale accuracy. Such SILs have been reported to achieve over three times enhancement in the extraction ratio, while SILs made of other materials such as GaAs can achieve over 10 times enhancement [33], due to this higher refractive index of GaAs compared to glass. The second advantage SILs could offer in our proposed photonic structure is the enhanced vertical confinement due to total internal reflection inside the cavity. By adding a slab of a similar refractive index to that of glass, FDTD simulations showed a four-fold enhancement in the cavity's Q-factor with the SIL in place. Combining our photonic crystal structure with a SIL could potentially enhance the extraction ratio of light by over 30 times. This makes combining a SIL with our proposed rod-type photonic crystal cavity a promising method for efficient extraction of light from 2D monolayer light emitters such as TMDCs.

## Conclusions

In this work we have designed and modelled a rod-type photonic crystal cavity to couple light from defects in 2D materials. By performing PWE numerical analysis on a pre-designed photonic crystal structure we have shown that a band-gap exists for TE-like modes. Subsequently, using 3D FDTD simulations and introducing a defect in the lattice, a localised cavity mode was found to have a Q-factor of 110. In this work we proposed transferring monolayers on top of a photonic crystal structure and allowing the dipping of the monolayer within the cavity to couple to the cavity mode. We compared the extraction ratio of light from a monolayer coupled to our proposed photonic structure with a case in which the monolayer is transferred onto a bulk substrate. We found that our structure can provide over three times enhancement in the extraction ratio. We performed a series of simulations to show how susceptible the design is to changes in the rod radius and the vertical position of the light emitter inside the cavity. Optimum enhancement was achieved for rods radius of $0.165\alpha$. The enhancement reduces to zero abruptly as $r$ is decreased but reduces gradually when $r$ is increased. The enhancement in the extracted light is also affected by the vertical position of the emitter within the cavity along the z-axis. Maximum enhancement is achieved when the dipole is at the centre of the cavity but it reduces dramatically as it reaches the top of the cavity. At the end, we discussed how solid immersion lenses placed on top of the cavity, can be used to enhance vertical confinement of the cavity mode within the vicinity of the cavity, increasing the total light extraction ratio to over 30 times. This enhancement is a strong step forward toward improved extraction efficiency of quantum and classical light from TMDC based devices.

## Acknowledgements

RJY acknowledges support by the Royal Society through a University Research Fellowship (UF110555). This work was also supported by grants from The Engineering and Physical Sciences Research Council in the UK.